\documentclass{ws-ijmpa}

\usepackage{amsmath}
\usepackage{graphicx}
\usepackage{dcolumn}
\usepackage{bm}
\usepackage{amssymb}
\usepackage{latexsym}

\def\setC{\mathbb{C}}

\def\setR{\mathbb{R}}

\newcommand{\lsim}{\lesssim}
\newcommand{\ie}{\textsl{i.e.~}}

\newcommand{\etal}{\textsl{et al.~}}

\newcommand{\mP}{m_{_{\mathrm Pl}}}
\newcommand{\GReCO}{${\cal G}\setR\varepsilon\setC{\cal O}$}

\begin{document}

\markboth{J\'er\^ome Martin}
{Inflation: where do we stand?}

%
\catchline{}{}{}{}{}
%

\title{INFLATION: WHERE DO WE STAND?}

\author{\footnotesize J\'ER\^OME MARTIN}

\address{Institut d'Astrophysique de Paris, \GReCO, FRE
2435-CNRS, 98bis boulevard Arago, \\ 75014 Paris, France.}

\maketitle

\pub{Received (Day Month Year)}{Revised (Day Month Year)}

\begin{abstract}
In this short review, the predictions of inflation are presented and
compared to the most recent measurements of the Cosmic Microwave
Background (CMB) anisotropy. It is argued that inflation is compatible
with these observations but that these ones are not yet accurate
enough to probe the details of the scenario.

\keywords{Cosmology; inflation; cosmological perturbations}
\end{abstract}

\section{Introduction}	

Inflation is presently our most convincing scenario of the very early
universe, not only because it is physically attractive, but also
because it makes a series of definite predictions that can be tested
concretely by means of astrophysical observations. In this short
article, we quickly review what are these predictions and describe how
the on-going high accuracy Cosmic Microwave Background (CMB)
anisotropy measurements can probe the physics of inflation. It is
argued that inflation is presently compatible with all the available
data but that these ones are not yet accurate enough to test the
details of the inflationary scenario. This review is organized as
follows. In the next section, we present the inflationary predictions
and, in particular, describe the theory of cosmological perturbations
of quantum-mechanical origin. In the next section, we compare these
predictions with the recently released CMB anisotropy
observations. Finally, we quickly present our conclusions.

\section{Basic Equations and Basic Predictions}

\subsection{Background Evolution and Cosmological Perturbations}

Inflation is a phase of the cosmic evolution which took place before
the hot Big Bang era~\cite{inflation}. This phase allows us to solve
many problems (horizon, flatness, monopoles etc \dots ) plaguing the
standard cosmological model. The main idea is that the evolution of
the Friedman-Lema\^{\i}tre-Robertson-Walker (FLRW) scale factor $a(t)$
was accelerated, ${\rm d}^2a/{\rm d}t^2>0$. This acceleration is
caused by a fluid the pressure of which is negative as required by the
Einstein equations of motion. Typically, since at very high energy a
fluid description of matter is likely to be inadequate, one uses
(quantum) field theory and models matter by one or many scalar
fields. Therefore, typically, the theory describing matter during
inflation possesses the Lagrangian~\cite{LR}
\begin{equation}
{\cal L}_{\rm matter}=-\frac12\sum _{i=1}^Ng^{\mu \nu}\partial _{\mu }
\varphi _i\partial _{\nu }\varphi _j -V(\varphi _1, \cdots, \varphi
_N)\, .
\end{equation}
At the end of the accelerated phase, the scalar field oscillates at
the bottom of its potential and decays into radiation. This is how one
smoothly connects inflation with the radiation dominated
era~\cite{reheating}.

\par

In order to go further, one must consider the theory of cosmological
perturbations of quantum-mechanical origin~\cite{MFB}. This theory
aims at explaining the origin and the evolution of the inhomogeneities
that are observed in our Universe. One of the most interesting aspects
of this theory is that it makes accurate predictions which are
currently under observational scrutiny. Technically, one uses the fact
that, in the very early universe, the inhomogeneities were small as
revealed by the COsmic Background Explorer (COBE) measurement of the
CMB anisotropy, $\delta T/T\simeq 10^{-5}$~\cite{cobe}. This permits
to work with a linear theory which is of course a great technical
simplification. Concretely, one writes~\cite{MFB} that the metric
tensor is given by
\begin{equation}
\label{metrictaylor}
\gamma _{\mu \nu }(\eta ,\vec{x})=\left[g_{\mu \nu}(\eta )+\epsilon
h_{\mu \nu }(\eta ,\vec{x}) +\epsilon ^2 \ell _{\mu \nu }(\eta
,\vec{x})+\cdots \right]{\rm d}x^{\mu } {\rm d}x^{\nu }\, ,
\end{equation}
where $g_{\mu \nu}(\eta )$ is the standard homogeneous and isotropic
FLRW metric. In the same manner, the matter sector can be expressed as
\begin{equation}
\label{fieldtaylor}
\varphi _i(\eta ,\vec{x})=\varphi _i(\eta )+\epsilon \delta \varphi_i
(\eta ,\vec{x}) +\epsilon ^2 \delta ^{(2)}\varphi _i(\eta ,\vec{x})
+\cdots \, ,
\end{equation}
where $i=1, \cdots , N$ with $N$ the total number of scalar fields. In
Eqs.~(\ref{metrictaylor}) and (\ref{fieldtaylor}), $\epsilon$ is a
small parameter in which the expansion is performed. The equations
linking the perturbed metric $h_{\mu \nu}$ and the perturbed scalar
fields $\delta \varphi_i $ are just the linearized Einstein
equations. 

\subsection{Density Perturbations}

Let us consider that two scalar fields, say $\varphi _1 $ and $\varphi
_2$, are present in the early universe (the following formalism can be
easily generalized to the case where there are more fields). We now
introduce the so-called perturbed adiabatic field and perturbed
entropy field defined by~\cite{noadia}
\begin{eqnarray}
\delta \sigma &=& \left(\cos \theta \right)\delta \varphi
_1+\left(\sin \theta \right)\delta \varphi_2\, , 
\quad 
\delta s = -\left( \sin \theta \right) \delta \varphi _1+\left(\cos
\theta \right) \delta \varphi _2\, ,
\end{eqnarray}
where 
\begin{equation}
\cos \theta = \frac{\varphi _1'}{\sqrt{\left(\varphi
_1'\right)^2+\left(\varphi _2'\right)^2}}\, ,\quad 
\sin \theta = \frac{\varphi _2'}{\sqrt{\left(\varphi
_1'\right)^2+\left(\varphi _2'\right)^2}}\, ,
\end{equation}
where a primes denotes a derivative with respect to the conformal
time.  The case where there is only one field corresponds to $\varphi
_2=0$ or $\theta =0$. In this case, it is obvious that there is no
entropy field, $\delta s=0$. Instead of working with $\delta \sigma $
and $\delta s$, it is more convenient to use $\delta s$ and the
generalized Mukhanov variable defined by
\begin{equation}
\label{defv}
v\equiv a\left[\delta \sigma +\frac{1}{{\cal H}}\sqrt{\left(\varphi
_1'\right)^2+\left(\varphi _2'\right)^2}\Phi \right] =\cos \theta
v_1+\sin \theta v_2\, ,
\end{equation}
where $\Phi $ is the Bardeen potential and $v_1$, $v_2$ the Mukhanov
variables associated to the first and second scalar fields
respectively. The quantity ${\cal H}$ is defined by $a'/a$, $a (\eta
)$ being the FLRW scale factor. Then, in the Fourier space, the
equations of motion of the system can be written as
\begin{eqnarray}
v''+v\left[k^2-\frac{\left(a\sqrt{\gamma }\right)''}{a\sqrt{\gamma
}}\right] &=& 2a\left(\theta '\delta s\right)'+4a\theta '\left({\cal
H}+\frac{\gamma '}{4\gamma }\right) \delta s\, , 
\\ 
\delta s''+2{\cal
H}\delta s'+\left(k^2+a^2V_{ss}+3\theta '^2\right)\delta s &=&
\frac{4\theta '}{\kappa \sqrt{\left(\varphi _1'\right)^2+\left(\varphi
_2'\right)^2}}k^2\Phi\, ,
\end{eqnarray}
where the function $\gamma $ is defined by $\gamma =1-{\cal H}'/{\cal
H}^2$ and the quantity $V_{ss}$ is given by $V_{s s}=\sin ^2\theta
V_{\varphi _1\varphi _1} -\sin 2\theta V_{\varphi _1\varphi _2}+\cos
^2\theta V_{\varphi _2\varphi _2}$. The quantity $k$ denotes the
wavenumber of the corresponding Fourier modes. We see that the system
can be reduced to equations describing parametric oscillators (the
time-dependent frequency coming from the ``interaction'' between the
perturbations and the background) together with an acting force term,
originating from the non-adiabatic nature of the fluctuations. The
integration of the above equations lead to the power spectra of the
adiabatic and non-adiabatic components during inflation (there is also
a mixed term). Each component is characterized by an amplitude and a
spectral index that can be evaluated for instance if the slow-roll
approximation holds~\cite{srmultifield}. In the single field case, the
spectrum is almost scale-invariant, the deviations from scale
invariance being small and given by the derivative of the inflaton
potential.

\par

The next question is now to propagate these spectra to the
radiation-domination era. For this purpose, let us introduce the
purely geometrical quantity $\zeta $ defined by~\cite{MS1}
\begin{equation}
\zeta =\Phi +\frac23\frac{{\cal H}^{-1}\Phi '+\Phi
}{1+\omega }\, ,
\end{equation}
where $\omega \equiv \rho /p$ is the equation of state parameter. One
can also work in terms of $\zeta _{_{\rm BST}}$ related to $\zeta $ by
\begin{equation}
\zeta _{_{\rm BST}}=-
\Phi -\frac23\frac{{\cal H}^{-1}\Phi '+\Phi
}{1+\omega }-\frac{k^2}{3\gamma {\cal H}^2}\Phi \, ,
\end{equation}
On super-Hubble scales, one has $\zeta _{_{\rm BST}}=-\zeta $. The
importance of the quantity $\zeta $ lies in the fact that, under
certain circumstances that we are going to discuss, it is conserved on
super-Hubble scales regardless of what happens during the complicated
phase where one goes from inflation to the radiation-dominated
era. Therefore, $\zeta$ can be viewed as a ``tracer'' for density
perturbations. Conservation of the perturbed stress-energy tensor
implies the following equation
\begin{equation}
\zeta _{_{\rm BST}}'=-\frac{{\cal H}}{\rho +p}\delta p_{\rm nad}
-\frac13 \partial _i\partial ^iv^{\rm (gi)}\, .
\end{equation}
On super-Hubble scales, the last term is negligible but the first one
is still important. This first term is the so-called non-adiabatic
pressure and is defined by the following expression: $\delta p_{\rm
nad}\equiv \delta p-c_{_{\rm S}}^2\delta \rho $, where $\delta \rho $,
$\delta p$ are the total energy density and pressure respectively and
$c_{_{\rm S}}^2\equiv p'/\rho '$ is the sound velocity. Expressing the
perturbed energy density and the perturbed pressure explicitly, one
has
\begin{eqnarray}
\delta p_{\rm nad} &=& \left(\delta p_1-c_{_{\rm S 1} }^2\delta \rho
_1\right) +\left(\delta p_2-c_{_{\rm S 2} }^2\delta \rho _2\right)
+\left(c_{_{\rm S1} }^2-c_{_{\rm S2} }^2\right) \frac{\left(\rho
_1+p_1\right)\left(\rho _2+p_2\right)}{\rho +p} S_{12}\, ,
\end{eqnarray}
where the quantity $S_{12}$ is given by
\begin{equation}
S_{12}=\frac{\delta \rho _1}{\rho _1+p_1}
-\frac{\delta \rho _2}{\rho _2+p_2}\, .
\end{equation}
We see that the non-adiabatic pressure contains two contributions. The
terms $\delta p_i-c_{_{\rm S i} }^2\delta \rho _i$ originate from
intrinsic entropy perturbations (if any) of the fluids under
consideration while the term proportional to $S_{12}$ represents the
entropy of mixing. The previous expressions are valid for any type of
matter. In the case of scalar fields, one obtains
\begin{equation}
\delta p_{\rm nad}=-\frac{2}{\kappa }\left(1-c_{_{\rm
S}}^2\right)\frac{k^2}{a^2}\Phi+\frac{2\theta '}{a^2}
\sqrt{\left(\varphi _1'\right)^2+\left(\varphi _2'\right)^2} \delta
s\, ,
\end{equation}
where $\kappa =8\pi /m_{_{\rm Pl}}^2$. On super-Hubble scales the
non-adiabatic pressure only comes from the entropy field $\delta s$,
\ie is sourced by the entropy of mixing and not by the intrinsic
entropy perturbations (which exist for a scalar field). 

\par

Therefore, if only one field is present, $\zeta $ is conserved and
this implies that the scale-invariant spectrum generated during
inflation is ``transferred'' to the perturbations of the various
components in the post-inflationary phase. Moreover, these
fluctuations are adiabatic, \ie they satisfy
\begin{equation}
\label{adiaic}
\frac{\delta \rho }{\rho }\biggl \vert _{\rm cdm}=
\frac{\delta \rho }{\rho }\biggl \vert _{\rm bayrons}=
\frac34 \frac{\delta \rho }{\rho }\biggl \vert _{\rm neutrinos}=
\frac34 \frac{\delta \rho }{\rho }\biggl \vert _{\rm photons}\, .
\end{equation}
These relations are then used to calculate observables like, for
instance, the multipole moments $C_{\ell }$ which characterize the CMB
anisotropies. 

\par

If more than one field is present the situation is more complicated
since one can no longer use the conservation of $\zeta $ to predict
the spectra in the post-inflationary phase (in addition, as already
mentioned above, there are several of them, ${\cal P}_{\zeta }$,
${\cal P}_{\delta s}$ and ${\cal P}_{\zeta-\delta s}$). Furthermore,
the relations~(\ref{adiaic}) are violated and, hence, the $C_{\ell }$
can strongly differ from the single field case.

\subsection{Gravitational Waves}

Gravitational waves are unavoidably produced in the early
universe~\cite{Gri}. Since the tensor modes do not couple to matter,
the gravitational wave spectrum is independent of the type of matter
(or of the number of scalar fields) present during inflation. It
reads~\cite{MS2}
\begin{equation}
{\cal P}_h=\frac{16H_{_{\rm inf}}^2}{\pi m_{_{\rm Pl}}^2}
\left[1-2\left(C+1\right)\epsilon -2\epsilon \ln
\frac{k}{k_*}\right] \, ,
\end{equation}
where $\epsilon =\mP ^2(V'/V)^2/(16\pi ^2)$ is the first slow-roll
parameter. It would be of utmost importance to detect the primordial
gravitational waves since this would provide a direct access to the
energy scale of inflation $H_{_{\rm inf}}$. Unfortunately, this is
observationally quite challenging.

\par

Another important check of the inflationary scenario would be to
measure the tensor spectrum to scalar spectrum ratio. Under quite
general circumstances, this can be expressed as~\cite{cccheck}
\begin{equation}
\label{ts}
\frac{T}{S}=-8n_{_{\rm T}}\sin ^2 \Delta \, ,
\end{equation}
where $n_{_{\rm T}}=-2\epsilon $ is the tensor spectral index and
$\Delta $ is a quantity which measures the correlation between
adiabatic and non-adiabatic perturbations. In absence of non-adiabatic
perturbations (\ie in the single field case), $\sin \Delta
=1$. Eq.~(\ref{ts}) tells us that tensor modes are sub-dominant since
the spectral index is expected to be small. A direct check of this
consistency relation would be a direct proof of inflation.

\subsection{Other Possibilities}

There are other effects that could modify the basic predictions of
inflation presented above. Here, we discuss two
possibilities. Firstly, there is the so-called trans-Planckian problem
of inflation~\cite{tpl}. It consists in the following. In a typical
model of inflation, the wavelength of the modes of astrophysical
interest today were, at the beginning of inflation, when the initial
conditions are chosen, smaller then the Planck length. In this regime,
the framework used to perform the calculations, \ie quantum field
theory in curved space-time, is likely to break down. In other words,
the predictions of inflation could be modified by short distance
physics. It is not easy to predict what these modifications could be
but a quite generic prediction is that superimposed oscillations in
the primordial power spectra should appear
\begin{equation}
\label{pssrs2}
{\cal P}_{\rm tpl} = {\cal P}_{\zeta } \Bigg\{ 1 - 2 |x|
\sigma_0\cos\left[ \frac{2 \epsilon }{\sigma_0} \ln
\left(\frac{k}{k_*} \right) + \psi \right] + \cdots  \Bigg\}\, ,
\end{equation}
where ${\cal P}_{\zeta }$ is the standard spectrum. The oscillations
in the power spectra are usually transferred to the multipole moments
$C_{\ell }$ which therefore also exhibit superimposed oscillations (of
course the amplitude and the frequency of these oscillations have
nothing to with the acoustic oscillations).

\par

A second interesting possibility is the presence of topological
defects that would be produced at the end of inflation (in the case
where the underlying model possesses several fields). This has
recently been studied in Ref.~\refcite{string} in the case of hybrid
inflation. Interestingly enough, it has been shown that, unless some
fine-tuning of the model parameters is present, the inflationary
multipole moments would be significantly changed.

\section{So Where Do We Stand?}

We now quickly discuss what are the consequences of the recently
released Wilkinson Microwave Anisotropy Probe (WMAP) data for
inflation~\cite{wmap}. First of all, these data are compatible with a
spatially flat universe, $\Omega _0 \simeq 1$. Secondly, the initial
(scalar) power spectrum is found to be compatible with scale
invariance but a small deviation from scale invariance cannot yet be
established with enough statistical confidence. No signal of
non-adiabatic fluctuations has been found. The first slow-roll
parameter $\epsilon $ is constrained to be $\epsilon \lsim 0.03$ and
we have an upper bound on the scale of inflation~\cite{saminf}
\begin{equation}
\frac{H_{_{\rm inf}}}{\mP}\lsim 1.4 \times 10^{-5}\, . 
\end{equation}
Some single field models are already ruled out, for instance the ones
with a potential of the form $V(\varphi )\propto \varphi ^p$, with $p
\ge 6$~\cite{saminf}. The case of the quartic potential $p=4$ is on
the border line while the massive potential $p=2$ is still
compatible~\cite{saminf}. Thirdly, no gravitational waves have been
detected. There is only an upper bound on the ratio $T/S$, namely
$T/S\lsim 0.3$~\cite{saminf}. Because there is no detection of
gravitational waves, the consistency check of inflation has obviously
not been verified. Fourthly, the statistical properties of the
fluctuations seem to be Gaussian which is compatible with single field
inflation. Recently, it has also been shown that there is a hint for
wiggles in the multipole moments but that the standard slow-roll model
remains the most probable one~\cite{wiggles}. In addition, this hint
is linked to the presence of the so-called ``cosmic variance
outliers'' which could very well disappeared with new data or be of
non-primordial origin (\ie linked to some astrophysical
foregrounds). Finally, no sign of topological defects has been
detected.

\par

As a general conclusion, one can say that the predictions of inflation
are compatible with the currently available data. Interestingly
enough, the data already allow us to exclude some models of
inflation. However, these data are not yet accurate enough to provide
us with something which could be considered as a definite proof of
inflation like a small deviation from scale invariance or, even better
but much more difficult, a detection of a background of stochastic
gravitational waves satisfying the consistency check of
inflation. Probably, we will have to wait for the next generation of
observations to achieve this ambitious goal.


\begin{thebibliography}{0}

\bibitem{inflation} A.~Guth, {\sl Phys.~Rev.}~{\bf D23}, 347 (1981);
A.~Linde, {\sl Phys.~Lett.}~{\bf B108}, 389 (1982); A.~Albrecht and
P.~J.~Steinhardt, {\sl Phys.~Rev.~Lett.}~{\bf 48}, 1220 (1982);
A.~Linde, {\sl Phys.~Lett.}~{\bf B129}, 177 (1983).

\bibitem{LR} D.~Lyth and A.~Riotto, {\sl Phys.~Rep.}~{\bf 314}, 1
(1999), {\tt hep-ph/9807278}.

\bibitem{reheating} M.~Turner, {\sl Phys.~Rev.}~{\bf D28}, 1243
(1983); L.~Kofman, A.~Linde and A.~Starobinsky, {\sl Phys.~Rev.}~{\bf
D56}, 3258 (1997), {\tt hep-ph/9704452}.

\bibitem{MFB} V.~F.~Mukhanov, H.~A.~Feldman, and R.~H.~Brandenberger,
{\sl Phys.~Rep.}~{\bf 215}, 203 (1992); J.~Martin, Proceedings of the
XXIV Brazilian National Meeting on Particles and Fields, Caxambu,
Brazil, (2004), {\tt astro-ph/0312492}; J.~Martin, Proceedings of the
40th Karpacz Winter School on Theoretical Physics, Poland, (2004),
{\tt hep-th/0406011}.

\bibitem{cobe} G.~F.~Smooth \etal, {\sl Astrophys.~J.}~{\bf 396}, L1
(1992).

\bibitem{noadia} C.~Gordon, D.~Wands, B.~A.~Basset and R.~Marteens, 
{\sl Phys.~Rev.}~{\bf D63}, 023506 (2001), {\tt astro-ph/0009131}.

\bibitem{srmultifield} N.~Bartolo, S.~Matarrese and A.~Riotto, {\sl
Phys.~Rev.}~{\bf D64}, 123504 (2001), {\tt astro-ph/0107502}.

\bibitem{MS1} D.~Lyth, {\sl Phys. Rev.}~{\bf D31}, 1792 (1984);
J.~Martin and D.~J.~Schwarz, {\sl Phys. Rev.}~{\bf D57}, 3302 (1998),
{\tt gr-qc/970449}.

\bibitem{Gri} L.~P.~Grishchuk, {\sl Zh.~Eksp.~Teor.~Fiz.}~{\bf 67},
825 (1974).

\bibitem{MS2} J.~Martin and D.~J.~Schwarz, {\sl Phys.~Rev.}~{\bf D62},
103520 (2000), {\tt astro-ph/9911225}; J.~Martin, A.~Riazuelo and
D.~J.~Schwarz, {\sl Astrophys.~J.}~{\bf 543}, L99 (2000), {\tt
astro-ph/0006392}.

\bibitem{cccheck} D.~Wands, N.~Bartolo, S.~Matarrese and A.~Riotto,
{\sl Phys.~Rev.}~{\bf D66}, 043520 (2002), {\tt astro-ph/0205253}.

\bibitem{tpl} J.~Martin and R.~H.~Brandenberger, {\sl Phys.~Rev.}~{\bf
D 63}, 123501 (2001), {\tt hep-th/0005209}; R.~H.~Brandenberger and
J.~Martin, {\sl Mod.~Phys.~Lett.}~{\bf A 16}, 999 (2001), {\tt
astro-ph/0005432}; J.~C.~Niemeyer, {\sl Phys.~Rev.}~{\bf D63}, 123502
(2001), {\tt hep-th/0005533}; M.~Lemoine, M.~Lubo, J.~Martin and
J.~P.~Uzan, {\sl Phys.~Rev.}~{\bf D 65}, 023510 (2002), {\tt
hep-th/0109128}; J.~Martin and R.~H.~Brandenberger, {\sl
Phys.~Rev.}~{\bf D 68}, 063513 (2003), {\tt hep-th/0305161}. J.~Martin
and R.~H.~Brandenberger, {\tt hep-th/0410223}.

\bibitem{string} J.~Rocher and M.~Sakellariadou, {\tt hep-ph/0405133}
and {\tt hep-ph/0406120}.

\bibitem{wmap} C.~L.~Bennet {\it et al.}, {\sl
Astrophys.~J.~Suppl.}~{\bf 148}, 1 (2003), {\tt astro-ph/0302207};
G.~Hinshaw {\it et al.}, {\sl Astrophys.~J.~Suppl.}~{\bf 148}, 135
(2003), {\tt astro-ph/0302217}; L.~Verde {\it et al.}, {\sl
Astrophys.~J.~Suppl.}~{\bf 148}, 195 (2003), {\tt astro-ph/0302218};
H.~V.~ Peiris {\it et al.}, {\sl Astrophys.~J.~Suppl.}~{\bf 148}, 213
(2003), {\tt astro-ph/0302225}.

\bibitem{saminf} S.~Leach and A.~Liddle, {\sl Phys.~Rev.}~{\bf D 68},
123508 (2003), {\tt astro-ph/0306305}.

\bibitem{wiggles} J.~Martin and C.~Ringeval, {\sl Phys.~Rev.}~{\bf D
69}, 083512 (2004), {\tt astro-ph/0310382}; J.~Martin and C.~Ringeval,
{\sl Phys.~Rev.}~{\bf D 69}, 127303 (2004), {\tt astro-ph/0402609};
J.~Martin and C.~Ringeval, {\tt hep-ph/0405249}.

\end{thebibliography}
\end{document}